# Upstream competition and exclusive content provision in media markets*

Kiho Yoon

Department of Economics, Korea University

145 Anam-ro, Seongbuk-gu, Seoul, Korea 02841

kiho@korea.ac.kr

https://kihoyoon.github.io

## Abstract

With a multilateral vertical contracting model of media markets, we examine upstream competition and contractual arrangements in content provision. We analyze the trade of content by the Nash bargaining solution and the downstream competition by the Hotelling location model. We characterize the equilibrium outcomes and the contractual arrangements for various vertical structures. We show that the possibility of exclusive contracts rises when the value of the premium content increases, the degree of horizontal differentiation in the downstream market decreases, the importance of advertising revenue decreases, and the relative bargaining power of upstream firm decreases.

* I thank an anonymous reviewer for helpful suggestions.

## 1. Introduction

The importance of content is ever-growing in the digital economy. Major sports events, blockbuster movies, and popular drama series are regarded as key elements for the success of video programming distributors. Many companies in the media and related industries spend huge sums of money to procure third parties' content as well as to produce their original content. For instance, Netflix spent $16.2 billion on content in 2024. Behind the recent merger and acquisition of big media companies, including Comcast-NBCUniversal, AT&T-Time Warner, and Disney-Fox, lies the desire to secure the premium content.

The vertical structure and the contractual arrangements in media markets are diverse. Many firms are vertically integrated and operating both in the production and in the distribution of content, whereas many other firms are only operating either in the production sector or in the distribution sector. The content is provided either by exclusive contracts or by non-exclusive contracts. For instance, DirecTV in the US holds the exclusive distribution rights for the National Football League (NFL) Sunday Ticket, and BSkyB in the UK has the first-run pay TV movie rights of Hollywood studios. On the other hand, several firms hold non-exclusive distribution rights for content such as Major League Baseball (MLB) Extra Innings and National Basketball Association (NBA) League Pass.

The purpose of this paper is to examine the upstream competition and the contractual arrangements in content provision. We construct a model with two upstream content providers (firm A and firm B) and two downstream platforms, i.e., content distributors (firm 1 and firm 2). The firms may be either vertically independent or vertically integrated. Each content provider bargains with each platform over the provision of its premium content. We analyze this bargaining process by the famous Nash bargaining solution. Each platform offers its basic content and any premium content it has acquired from the content providers to the final consumers and earns the subscription revenue. The platforms are horizontally differentiated, and we analyze the downstream competition for subscribers by the Hotelling location model as it is widely adopted in the literature on the media industry.[1]

We first study the case when the firms are vertically independent. We show that the possibility of exclusive contracts rises when the value of the premium content increases, the degree of horizontal differentiation in the downstream market decreases, and the importance of upstream advertising revenue decreases. This is because (i) the platforms have a strong incentive to exclusively secure the premium content and gain a competitive advantage when the value of the premium content is high and/or the downstream competition is intense, and (ii) the opportunity cost of exclusive contracts for the content providers is small when the advertising revenue is low. We also show that the possibility that the content providers offer exclusive contract to the same platform rather than to different platforms rises when the value of the premium content increases and the importance of upstream advertising revenue decreases. Besides, we show that the possibility of exclusive contracts rises when the relative bargaining power of the upstream firm gets weaker, and moreover, the content providers offer exclusive contract to different platforms when the relative bargaining power is sufficiently weak.

[1] See, for instance, Gabszewicz *et al.* (2001, 2002, 2004), Gal-Or and Dukes (2003), Anderson and Coate (2005), Peitz and Valleti (2008), Stennek (2014), Weeds (2014, 2016), and D'Annunzio (2017).

We next study the case when there are two vertical integrations: firm A and firm 1 are vertically integrated as well as firm B and firm 2 are vertically integrated. We find that the previous comparative static results continue to hold: The possibility of exclusive contracts rises when the value of the premium content increases, the degree of horizontal differentiation in the downstream market decreases, the importance of advertising revenue decreases, and the relative bargaining power of the upstream firm gets weaker.

We also study the case when there is one vertical integration, i.e., one content provider and one platform, say firm A and firm 1, are vertically integrated while firm B and firm 2 are independent. This case is quite complicated and we cannot obtain unambiguous comparative static results. However, when there is no upstream advertising revenue, we find that there are only three possible contractual arrangements depending on the values of the parameters.

We turn to the literature review. Armstrong (1999) showed that a monopolistic and vertically independent content provider offers an exclusive contract when lump-sum fees are used for the sale of content. Harbord and Ottaviani (2001), on the other hand, showed that the content provider may also offer a non-exclusive contract when per-subscriber wholesale fees for content are used. Weeds (2014) showed that the content provider may choose exclusive or non-exclusive provision when the platforms are advertising-funded instead of subscription-funded. Weeds (2016) studied a model with a vertically integrated firm and an independent platform. She showed that, in the static model, the integrated firm always supplies its premium content to the downstream rival. In contrast, in a dynamic model with switching costs, the integrated firm may choose exclusivity if the value of its premium content is high and/or the degree of horizontal differentiation in the downstream market is low. D'Annunzio (2017) showed that a vertically integrated firm invests less in content than a vertically independent content provider. Stennek (2014) set up a model with a monopolistic and vertically independent content provider and two platforms, and showed that exclusivity leads to higher investment in content. In contrast to all the other papers cited above which assumed that the content provider makes a take-it-or-leave-it offer, that is, assumed that the content provider holds all the bargaining power, the last paper applied Rubinstein's alternating-offer bargaining process to the determination of content payments.

Compared to the previous literature, the novel features of the current paper are (i) to consider the upstream competition in media market,[2] and (ii) to examine the effect of relative

---

[2] D'Annunzio (2017) briefly considers two content providers in section 5.5. However, the upstream firms provide perfectly substitutable premium content and each platform can acquire at most one upstream firm's premium content. In contrast, the upstream firms' premium content can be either substitutes or complements,

bargaining power on contractual form. In particular, we allow the relative bargaining power to take any value between zero and one.[3] This is in accordance with industry practice: Firms bargain in media markets if for no other reason than that both the upstream and downstream firms are large and hence have market power. The US FTC explicitly adopted the bargaining framework in assessing the competitive effects of the Comcast-NBCUniversal vertical merger case.[4]

The approach we take is similar to the "Nash-in-Nash" approach in the literature. Beginning with Horn and Wolinsky (1988), this literature studies the settings with many upstream firms and many downstream firms and characterizes the Nash equilibria of the contracting game in which the outcome in each pair of an upstream firm and a downstream firm is given by the Nash bargaining solution. Representative works include Dobson and Waterson (2007), Rey and Vergé (2019), and Collard-Wexler *et al.* (2019).[5] In particular, Ho and Lee (2019), Ghili (2022), and Liebman (2022) incorporate the threat of replacement into the Nash-in-Nash approach.[6] This literature, however, does not consider the exclusive contract as an explicit strategy.

This paper belongs to the broad literature of vertical relations and vertical contracting. Representative works include Hart and Tirole (1990), O'Brien and Shaffer (1992), McAfee and Schwartz (1994), Dobson and Waterson (1996), and Rey and Vergé (2004). In particular, Nocke and Rey (2018) presented a general model of interlocking vertical relationships with such features as the absence of any restriction on contracts, secret contracting, balanced bargaining power, and Cournot competition in the downstream market. They showed that pairwise exclusivity, i.e., upstream firms offer exclusive contracts to different downstream firms, is the equilibrium outcome. In contrast, reflecting the reality of actual media markets, we assume that downstream platforms are horizontally differentiated and show that diverse outcomes emerge depending on the values of the parameter.

The plan of the paper is as follows. Sections 2-4 sequentially analyze the cases of vertical separation, two vertical integrations, and one vertical integration. Section 5 compares these vertical structures and briefly examines firms' incentives to vertically integrate as well as the welfare implications. Section 6 concludes with a discussion on the limitations of the model and possible extensions.

---

and each platform can acquire zero, one, or two upstream firms' premium content in our model.

[3] Stennek (2014) considers Rubinstein's alternating-offer bargaining, whose outcome is the same as the Nash bargaining solution when the bargaining parties have equal bargaining power.

[4] See Rogerson (2012) for a detailed discussion.

[5] See also footnotes 4 and 5 in Collard-Wexler *et al.* (2019) as well as footnote 16 in Nocke and Rey (2018).

[6] See the next section for further discussion.

## 2. Vertically independent firms

There are two upstream content providers and two downstream platforms. The upstream firms can be national broadcast networks, movie studios, or sports content providers. The downstream firms can be broadcast stations, multichannel video programming distributors (MVPDs), or online video distributors (OVDs).[7] The upstream firms are denoted by firm A and firm B, and the downstream firms are denoted by firm 1 and firm 2.

The profit of a content provider consists of the lump-sum fees it receives from the platforms for its content and the advertising revenue minus any costs. We assume for simplicity that the marginal cost of the content provider is zero.[8] We also assume that the content production cost is already sunk. The profit of a platform is the subscription charges it receives from final consumers minus its possible payments made to the content providers. We assume that the marginal cost of the platform is zero.[9] We also assume that the platforms do not earn any advertising revenue.[10]

Each platform initially owns its own basic content, and each content provider may provide its own premium content to the platforms. The order of moves is as follows. First, each content provider and each platform bargain over the lump-sum fee for the provision of content. Second, the platforms compete to attract subscribers. Each platform offers its own basic content and the premium content it may have acquired in the bargaining stage from the content providers to final consumers in return for the subscription charges. By backward induction, we analyze the downstream market for subscribers and then move on to the bargaining for content provision.

We note that the sale of the premium content on the basis of lump-sum fees is industry practice. We also note that Rey and Vergé (2019) have shown in a model of multilateral vertical contracting that equilibrium tariffs are cost-based, which turn out to be lump-sum fees when the marginal cost is zero. Observe that there does not exist double marginalization when the payment for the content is lump-sum.

---

[7] Online video distributors include Netflix, Amazon, Hulu, and Disney+. They are also known as the over-the-top (OTT) service providers.

[8] The case of positive marginal cost does not lead to any qualitatively different result.

[9] The case of positive marginal cost does not lead to any qualitatively different result since it is only a normalization to set the marginal cost to zero.

[10] It is possible that the results are qualitatively different when the platforms earn advertising revenue. We leave it to future research agenda. Weeds (2014) contains a relevant analysis for the case when there is one monopolistic upstream firm.

## 2-1. The downstream market: competition for subscribers

The competition between the platforms for subscribers is analyzed by the Hotelling location model. Consumers of mass one are uniformly distributed over the unit interval [0,1]. Firm 1 is located at the left end (zero) and firm 2 is located at the right end (one). Each consumer's gross utility when subscribing to platform $d = 1,2$ is $v_d$, where

$$v_d = \begin{cases} v, & \text{when } d \text{ offers only basic content;} \\ v + \alpha, & \text{when } d \text{ also offers one firm's premium content;} \\ v + \alpha + \beta, & \text{when } d \text{ also offers two firms' premium content.} \end{cases}$$

Thus, each consumer's gross utility is $v$ when platform $d$ offers only its own basic content; $v + \alpha$ when platform $d$ offers the premium content of one upstream firm, be it A or B, in addition to its basic content; $v + \alpha + \beta$ when platform $d$ offers the premium content of both upstream firms in addition to its own basic content. We have $\alpha \geq 0$ and $\beta \geq 0$. We may interpret that the premium content packages of the upstream firms are independent when $\alpha = \beta$; substitutes when $\alpha > \beta$; and complements when $\alpha < \beta$, if we consider the incremental value of additional content. For instance, they are perfect substitutes when $\beta = 0$ and perfect complements when $\alpha = 0$.

Let $p_d$ denote the subscription charge of platform $d$. The net utility of consumer $x \in [0,1]$ is $v_1 - tx - p_1$ when subscribing to firm 1 and $v_2 - t(1 - x) - p_2$ when subscribing to firm 2. Note that the parameter $t$, often termed as the 'transport cost,' measures the degree of horizontal differentiation between firm 1 and firm 2.

The equilibrium price $p_d$, quantity $q_d$, and profit $\pi_d$ in the downstream market are given as

$$p_d = t + \frac{v_d - v_{d'}}{3}; q_d = \frac{1}{2} + \frac{v_d - v_{d'}}{6t}; \pi_d = \frac{1}{2t}\left(t + \frac{v_d - v_{d'}}{3}\right)^2,$$

where $d = 1,2$ and $d'$ is the downstream competitor such that $d + d' = 3$. To maintain the competition in the downstream market, i.e., to rule out the situation when one of the downstream firms exits the market, we assume that

$$\alpha + \beta < 3t.$$

## 2-2. The upstream market: bargaining and exclusive contracts

In this subsection, we analyze the bargaining process between the content providers and the platforms. To assist the readers' understanding and to present the results in a more intuitive fashion, we first perform a preliminary analysis with a simplified setup and then take on the general analysis.

### 2-2-1. A preliminary analysis

We assume for the preliminary analysis that (i) there is no advertising revenue and that (ii) the upstream firm holds all the bargaining power in the negotiation for its premium content. Thus, firm A makes a take-it-or-leave-it offer in the negotiation with firm 1 for its premium content, and the same holds true in the negotiations between firm A and firm 2, between firm B and firm 1, and between firm B and firm 2. Each content provider independently decides whether to provide its premium content exclusively to one of the downstream firms or non-exclusively to both downstream firms.

This market for the premium content can be analyzed by a strategic form game. Each content provider has three strategies in this game: the exclusive contract to firm 1 (denoted by E1), the exclusive contract to firm 2 (denoted by E2), and the non-exclusive contract (denoted by N). The game matrix is given as follows, with firm A being the row player and firm B being the column player.

| | E1 | E2 | N |
|---|---|---|---|
| E1 | $\frac{(\alpha+\beta)(6t+\alpha+\beta)}{18t}$, $\frac{(\alpha+\beta)(6t+\alpha+\beta)}{18t}$ | $\frac{(\alpha+\beta)(6t-\alpha-\beta)}{18t}$, $\frac{(\alpha+\beta)(6t-\alpha-\beta)}{18t}$ | $\frac{2\beta}{3}$, $\frac{6(\alpha+\beta)t+\beta^2-\alpha^2-2\alpha\beta}{18t}$ |
| E2 | $\frac{(\alpha+\beta)(6t-\alpha-\beta)}{18t}$, $\frac{(\alpha+\beta)(6t-\alpha-\beta)}{18t}$ | $\frac{(\alpha+\beta)(6t+\alpha+\beta)}{18t}$, $\frac{(\alpha+\beta)(6t+\alpha+\beta)}{18t}$ | $\frac{2\beta}{3}$, $\frac{6(\alpha+\beta)t+\beta^2-\alpha^2-2\alpha\beta}{18t}$ |
| N | $\frac{6(\alpha+\beta)t+\beta^2-\alpha^2-2\alpha\beta}{18t}$, $\frac{2\beta}{3}$ | $\frac{6(\alpha+\beta)t+\beta^2-\alpha^2-2\alpha\beta}{18t}$, $\frac{2\beta}{3}$ | $\frac{2\beta}{3}-\frac{\beta^2}{9t}$, $\frac{2\beta}{3}-\frac{\beta^2}{9t}$ |

To understand the payoff, let us first consider the strategy profile (E1, E1). Firm A receives a lump-sum fee for its premium content of

$$\frac{1}{2t}\left(t+\frac{\alpha+\beta}{3}\right)^2-\frac{t}{2}=\frac{(\alpha+\beta)(6t+\alpha+\beta)}{18t}$$

by offering the exclusive contract to firm 1 to which firm B also offers the exclusive contract. Observe that (i) if firm 1 accepts firm A's offer, then its profit is $(t+(\alpha+\beta)/3)^2/(2t)$ since it secures the premium content of both upstream firms while the other platform does not have any premium content, (ii) if firm 1 refuses firm A's offer, then its profit is $t/2$ since firm A will offer the exclusive contract to firm 2 instead and so each platform now secures the premium content of one upstream firm, and so (iii) firm A can obtain the difference in profits by making a take-it-or-leave-it offer since it has all the bargaining power in the negotiation with the downstream firm.[11] We note that the outcome when firm A offers the exclusive contract to the other platform is the disagreement outcome of the negotiation between firm A and the platform to which firm B offers the exclusive contract. We also note that the other firm will accept firm A's exclusive offer (with paying the lump-sum fee) when the platform to which firm B offers the exclusive contract refuses firm A's offer.

Observe that this reasoning is similar in spirit to that of "Nash-in-Nash with threat of replacement (NNTR)" approach, as proposed by Ho and Lee (2019), Ghili (2022), and Liebman (2022) in the study of medical insurance markets. While the traditional Nash-in-Nash approach stipulates that what happens between firm A and firm 1 does not affect what happens between firm A and firm 2, this new approach explicitly considers the threat that firm A would replace firm 1 with firm 2 in case the bargaining breaks down. Similar logic holds in all of the cases below.

To see the reasoning just one more time, consider the strategy profile (N, E1). Firm A receives a lump-sum fee of

$$\frac{1}{2t}\left(t+\frac{\beta}{3}\right)^2-\frac{t}{2}$$

from the platform to which firm B offers the exclusive contract and receives a lump-sum fee of

$$\frac{1}{2t}\left(t-\frac{\beta}{3}\right)^2-\frac{1}{2t}\left(t-\frac{\alpha+\beta}{3}\right)^2$$

from the other platform, thus receiving a total lump-sum of

$$\frac{6(\alpha+\beta)t+\beta^2-\alpha^2-2\alpha\beta}{18t}.$$

In the negotiation between firm A and firm 1, (i) if firm 1 accepts firm A's offer, then its profit is $(t+\beta/3)^2/(2t)$ since it secures the premium content of both upstream firms while the other platform secures the premium content of one upstream firm, and (ii) if firm 1 refuses firm A's offer, then its profit is $t/2$ since each platform now secures the premium content of one upstream provider, thus (iii) firm A can obtain the difference in profits by making a take-it-or-leave-it offer. Similar reasoning applies to the negotiation between firm A and firm 2. We note that during a negotiation both parties assume all other negotiations will succeed.

Observe that it is better for an upstream firm to offer the exclusive contract to the same platform to which the other upstream firm offers the exclusive contract rather than to the

[11] Observe that the platform's *net* profit net of the payments to the content providers when it secures the premium content of both upstream firms is $(t+(\alpha+\beta)/3)^2/(2\mathrm{t})-2(\alpha+\beta)(6t+\alpha+\beta)/(18t)$. This must be nonnegative if no loss condition is imposed. This condition is equivalent to the condition that $3\mathrm{t} \geq (1+\sqrt{2})(\alpha+\beta)$, and we can assume, if needed, that this inequality holds.

other platform.[12] Observe also that

$$\frac{(\alpha+\beta)(6t+\alpha+\beta)}{18t}-\frac{6(\alpha+\beta)t+\beta^2-\alpha^2-2\alpha\beta}{18t}=\frac{\alpha(\alpha+2\beta)}{9t}>0$$

holds. Thus, the equilibria are (E1, E1) and (E2, E2), that is, the upstream firms offer exclusive contracts to the same platform.

We want to discuss the assumption of symmetric upstream content providers at this juncture. What might happen if we introduce asymmetry such that each consumer's gross utility $v_d$ when subscribing to platform $d = 1,2$ is given as (i) $v$ when $d$ offers only basic content; (ii) $v + \alpha_u$ when $d$ also offers upstream firm $u$'s premium content for $u = A, B$; and (iii) $v + \gamma$ when $d$ also offers both upstream firms' premium content, with $\gamma \geq \alpha_A \geq \alpha_B$? It can be checked that, while firms' profits may differ, the equilibrium structure does not change. That is, the exclusive contracts to the same platform, (E1, E1) and (E2, E2), remain the only equilibria. The reason is essentially that firms' incentives are qualitatively unaltered by asymmetry.

### 2-2-2. The general analysis

Let us return to the general setup and assume that the content providers earn advertising revenue in addition to lump-sum fees for their content. Each content provider earns advertising revenue of $r$ per consumer to whom its premium content is reached. Hence, a content provider earns the total advertising revenue of $rq$ when the mass of consumers who consume its premium content is $q$.

In the negotiation between an upstream firm and a downstream firm, we define the parameter $\lambda \in [0,1]$ to represent the relative bargaining power of the upstream firm. When $\lambda = 1$, the upstream firm holds all the bargaining power and makes a take-it-or-leave-it offer to the downstream firm. When $\lambda = 0$, on the contrary, the downstream firm holds all the bargaining power and makes a take-it-or-leave-it offer to the upstream firm. When $\lambda = 1/2$, the upstream firm and the downstream firm hold equal bargaining power. We note that the parameter $\lambda$ may reflect the firms' discount rates, degree of risk aversion, and so on.

The equilibrium concept we employ for the negotiation process is the well-known Nash bargaining solution. Let $b_U$ and $b_D$ represents, respectively, the payoff that the upstream firm and the downstream firm, respectively, gets when the negotiation succeeds, and let $n_U$ and $n_D$ represents, respectively, the payoff that the upstream firm and the downstream firm, respectively, gets when the negotiation fails. That is, $n_U$ and $n_D$ are the payoffs from the disagreement outcome, or alternatively, the threat point. Let $l$ denote the lump-sum fee that the upstream firm receives from the downstream firm for its premium content. Then, the Nash bargaining solution is the value of $l$ that maximizes

$$(b_U + l - n_U)^{\lambda}(b_D - l - n_D)^{1-\lambda},$$

and the solution is

$$l = \lambda(b_D - n_D) - (1-\lambda)(b_U - n_U).$$

[12] This result holds since the competitive pressure is reduced when firms are more vertically differentiated, so the incremental profit of the firm with a higher market share is bigger than the loss of the firm with a lower market share. This feature is common to models of product differentiation with linear demands under Bertrand and Cournot competition. See Bester and Petrakis (1993).

Thus, we may imagine that the upstream firm receives the fraction $\lambda$ of the gain of the downstream firm, $(b_D - n_D)$, but pays the fraction $1-\lambda$ of the gain of the upstream firm, $(b_U - n_U)$.

Since the firms are symmetric, it is analytically convenient to treat the exclusive contract to firm 1 and the exclusive contract to firm 2 at once. Hence, we let the strategy E denote the exclusive contract offered either to firm 1 or to firm 2.

(1) When firm B chooses strategy E:

(1-1) When firm A chooses strategy E:

If firm A chooses strategy E, then it can offer the exclusive contract (i) to the same platform to which firm B offers the exclusive contract or (ii) to the other platform than B offers the exclusive contract. Assume without loss of generality that firm B offers the exclusive contract to firm 1. In the former case, with the disagreement outcome of A's providing its premium content exclusively to 2, the gain of 1 is

$$\frac{1}{2t}\left(t + \frac{\alpha+\beta}{3}\right)^2 - \frac{t}{2}$$

and the gain of A is

$$\left(\frac{1}{2} + \frac{\alpha+\beta}{6t}\right)r - \frac{r}{2} = \frac{\alpha+\beta}{6t}r.$$

Hence,

$$l = \lambda\left(\frac{1}{2t}\left(t + \frac{\alpha+\beta}{3}\right)^2 - \frac{t}{2}\right) - (1-\lambda)\frac{\alpha+\beta}{6t}r,$$

$$\pi_A = \left(\frac{1}{2} + \frac{\alpha+\beta}{6t}\right)r + l = \frac{\lambda(\alpha+\beta)(6t+3r+\alpha+\beta)}{18t} + \frac{r}{2} \quad . \qquad (1)$$

In the latter case, with the disagreement outcome of A's providing its premium content exclusively to 1, the gain of 2 is

$$\frac{t}{2} - \frac{1}{2t}\left(t - \frac{\alpha + \beta}{3}\right)^2$$

and the gain of A is

$$\frac{r}{2} - \left(\frac{1}{2} + \frac{\alpha+\beta}{6t}\right) r = -\frac{\alpha+\beta}{6t} r.$$

Hence,

$$l = \lambda\left(\frac{t}{2} - \frac{1}{2t}\left(t - \frac{\alpha + \beta}{3}\right)^2\right) - (1 - \lambda)\left(-\frac{\alpha + \beta}{6t} r\right),$$

$$\pi_A = \frac{r}{2} + l = \frac{\lambda(\alpha+\beta)(6t-3r-\alpha-\beta)}{18t} + \left(\frac{1}{2} + \frac{\alpha+\beta}{6t}\right) r. \quad (2)$$

Comparing (1) and (2), firm A offers the exclusive contract to the same platform to which firm B offers the exclusive contract when

$$\lambda > \tilde{\lambda} \equiv \frac{3r}{2\alpha+2\beta+6r} \ ,$$

offers the exclusive contract to the other platform when $\lambda < \tilde{\lambda}$, and is indifferent between these two alternatives when $\lambda = \tilde{\lambda}$.

Observe that $\tilde{\lambda} < 1/2$, and thus firm A offers the exclusive contract to the same platform to which firm B offers the exclusive contract when the upstream firm and the downstream firm hold equal bargaining power. We also note that $\tilde{\lambda}$ increases when $\alpha$ decreases, $\beta$ decreases, and $r$ increases. Hence, the possibility that firm A offers the exclusive contract to the same platform to which firm B offers the exclusive contract rises when $\alpha$ increases, $\beta$ increases, and $r$ decreases.

(1-2) When firm A chooses strategy N:

In the negotiation between A and 1, with the disagreement outcome of A's providing its premium content only to 2, the gain of 1 is

$$\frac{1}{2t}\left(t + \frac{\beta}{3}\right)^2 - \frac{t}{2}$$

and the gain of A is

$$\left(1 - \frac{1}{2} r\right) = \frac{r}{2} \ .$$

Hence,

$$l_{A,1} = \lambda\left(\frac{1}{2t}\left(t + \frac{\beta}{3}\right)^2 - \frac{t}{2}\right) - (1 - \lambda)\frac{r}{2} \ .$$

In the negotiation between A and 2, with the disagreement outcome of A's providing its premium content only to 1, the gain of 2 is

$$\frac{1}{2t}\left(t-\frac{\beta}{3}\right)^2-\frac{1}{2t}\left(t-\frac{\alpha+\beta}{3}\right)^2$$

and the gain of A is

$$\left(1-\frac{1}{2}-\frac{\alpha+\beta}{6t}\right)r=\left(\frac{1}{2}-\frac{\alpha+\beta}{6t}\right)r.$$

Hence,

$$l_{A,2}=\left(\frac{1}{2t}\left(t-\frac{\beta}{3}\right)^2-\frac{1}{2t}\left(t-\frac{\alpha+\beta}{3}\right)^2\right)-(1-\lambda)\left(\frac{1}{2}-\frac{\alpha+\beta}{6t}\right)r,$$

$$\pi_A=r+l_{A,1}+l_{A,2}=\lambda\frac{6(\alpha+\beta)t+\beta^2-\alpha^2-2\alpha\beta+3(6t-\alpha-\beta)r}{18t}+\frac{(\alpha+\beta)r}{6t}. \qquad (3)$$

(1-3) The optimal choice of firm A:

Consider first the case when $\lambda<\tilde{\lambda}$ holds. Then, the profit of (2) is higher than the profit of (1). Subtracting the profit of (3) from the profit of (2), we get

$$\frac{r}{2}-\left(r+\frac{\beta^2}{9t}\right)\lambda,$$

which is decreasing in $\lambda$ and equal to

$$\frac{r(3t(\alpha+\beta)-\beta^2)}{6t(\alpha+\beta+3r)}$$

when $\lambda=\tilde{\lambda}$. Since the last expression is positive by our assumption that $\alpha+\beta<3t$, we conclude that firm A offers the exclusive contract to the other downstream firm than firm B offers the exclusive contract. Let us denote this strategy by E(o).

Consider next the case when $\lambda>\tilde{\lambda}$ holds. Then, the profit of (1) is higher than the profit of (2). Subtracting the profit of (3) from the profit of (1), we get

$$\frac{2\big(\alpha(\alpha+2\beta)-3r(3t-\alpha-\beta)\big)\lambda+3r(3t-\alpha-\beta)}{18t}.$$

If $\alpha(\alpha+2\beta)-3r(3t-\alpha-\beta)\geq 0$, then the expression above is non-negative and thus firm A offers the exclusive contract to the same platform to which firm B offers the exclusive contract.[13] Let us denote this strategy by E(s). If $\alpha(\alpha+2\beta)-3r(3t-\alpha-\beta)<0$, then firm A chooses E(s) when

$$\lambda < \hat{\lambda} \equiv \frac{3r(3t-\alpha-\beta)}{6r(3t-\alpha-\beta)-2\alpha(\alpha+2\beta)}$$

and N when $\lambda > \hat{\lambda}$. Observe that $\hat{\lambda} > 1/2$ and so $\hat{\lambda} > \tilde{\lambda}$. This in particular implies that firm A offers the exclusive contract to the same platform to which firm B offers the exclusive contract, i.e., firm A chooses E(s), when the upstream firm and the downstream firm hold equal bargaining power. We also note that $\hat{\lambda}$ increases when $\alpha$ increases, $\beta$ increases, $t$ decreases, and $r$ decreases. Hence, the possibility that firm A offers the exclusive contract rises when $\alpha$ increases, $\beta$ increases, $t$ decreases, and $r$ decreases.

---

[13] This expression is equal to zero when $r=0$ and $\lambda=0$. Hence, firm A is indifferent between E(s) and N, and we assume that firm A chooses E(s) in this case.

Summarizing the discussion,

**Lemma 1.** *The optimal choice of firm A when firm B chooses E is as follows.*

*(i)* *When* $\lambda < \tilde{\lambda}$*: firm A chooses E(o).*

*(ii)* *When* $\lambda > \tilde{\lambda}$ *and* $\alpha(\alpha+2\beta) - 3r(3t-\alpha-\beta) \geq 0$*: firm A chooses E(s).*

*(iii)* *When* $\lambda > \tilde{\lambda}$ *and* $\alpha(\alpha+2\beta) - 3r(3t-\alpha-\beta) < 0$*: firm A chooses E(s) if* $\lambda < \hat{\lambda}$*, and chooses N if* $\lambda > \hat{\lambda}$*.*

(2) When firm B chooses strategy N:

(2-1) When firm A chooses strategy E:

With the disagreement outcome of A's providing its premium content to the other platform, the gain of the downstream firm is

$$\frac{1}{2t}\left(t+\frac{\beta}{3}\right)^2 - \frac{1}{2t}\left(t-\frac{\beta}{3}\right)^2 = \frac{2\beta}{3}$$

and the gain of A is zero. Hence,

$$l = \lambda\frac{2\beta}{3},$$

$$\pi_A = \lambda\frac{2\beta}{3} + \left(\frac{1}{2}+\frac{\beta}{6t}\right)r. \quad (4)$$

(2-2) When firm A chooses strategy N:

In the negotiation with each of the downstream firms, with the disagreement outcome of A's providing its premium content only to the other downstream firm, the gain of the downstream firm is

$$\frac{t}{2} - \frac{1}{2t}\left(t-\frac{\beta}{3}\right)^2 = \frac{\beta}{3} - \frac{\beta^2}{18t}$$

and the gain of A is

$$\left(1-\left(\frac{1}{2}+\frac{\beta}{6t}\right)\right)r=\left(\frac{1}{2}-\frac{\beta}{6t}\right)r.$$

Hence,

$$\pi_A = r + 2\left[\lambda\left(\frac{\beta}{3}-\frac{\beta^2}{18t}\right)-(1-\lambda)\left(\frac{1}{2}-\frac{\beta}{6t}\right)r\right] = \lambda\left(\frac{2\beta}{3}-\frac{\beta^2}{9t}+\left(1-\frac{\beta}{3t}\right)r\right)+\frac{\beta r}{3t}. \quad (5)$$

(2-3) The optimal choice of firm A:

Subtracting the profit of (5) from the profit of (4), we get

$$\frac{2\left(\beta^2-3r(3t-\beta)\right)\lambda+3r(3t-\beta)}{18t}.$$

If $\beta^2-3r(3t-\beta)\geq 0$, then the expression above is non-negative and thus firm A offers the exclusive contract.[14] If $\beta^2-3r(3t-\beta)<0$, then firm A chooses E when

$$\lambda<\bar{\lambda}\equiv\frac{3r(3t-\beta)}{6r(3t-\beta)-2\beta^2}$$

and N when $\lambda>\bar{\lambda}$. Observe that $\bar{\lambda}>1/2$, and thus firm A offers the exclusive contract to one of the platforms when the upstream firm and the downstream firm hold equal bargaining power. We also note that $\bar{\lambda}$ increases when $\beta$ increases, $t$ decreases, and $r$ decreases. Hence, the possibility that firm A offers the exclusive contract rises when $\beta$ increases, $t$ decreases, and $r$ decreases.

Summarizing the discussion,

**Lemma 2.** *The optimal choice of firm A when firm B chooses N is as follows.*

*(i)* *When* $\beta^2-3r(3t-\beta)\geq 0$*: firm A chooses E.*

*(ii)* *When* $\beta^2-3r(3t-\beta)<0$ *and* $\lambda<\bar{\lambda}$*: firm A chooses E.*

*(iii)* *When* $\beta^2-3r(3t-\beta)<0$ *and* $\lambda>\bar{\lambda}$*: firm A chooses N.*

(3) The equilibrium analysis

Recall that $\tilde{\lambda}<1/2$, $\hat{\lambda}>1/2$, and $\bar{\lambda}>1/2$. This in particular implies that $\tilde{\lambda}<\hat{\lambda}$ and $\tilde{\lambda}<\bar{\lambda}$. Moreover, we have

[14] This expression is equal to zero when $r=0$ and $\lambda=0$. Hence, firm A is indifferent between E and N, and we assume that firm A chooses E in this case.

$$\hat{\lambda} - \bar{\lambda} = \frac{3r}{2}\left(\frac{3t-\beta}{3r(3t-\beta)-\beta^2} + \frac{3t-\alpha-\beta}{3r(3t-\alpha-\beta)-\alpha(\alpha+2\beta)}\right) > 0$$

and thus

$$\tilde{\lambda} < \bar{\lambda} < \hat{\lambda}$$

when both $\alpha(\alpha+2\beta) - 3r(3t-\alpha-\beta) < 0$ and $\beta^2 - 3r(3t-\beta) < 0$ hold.

Let us assume for the time being that $\alpha(\alpha+2\beta) - 3r(3t-\alpha-\beta) \geq 0$ holds when $\beta^2 - 3r(3t-\beta) \geq 0$ holds. (We also discuss the case when $\beta^2 - 3r(3t-\beta) \geq 0$ but $\alpha(\alpha+2\beta) - 3r(3t-\alpha-\beta) < 0$ below.) Note that this relationship is true if $\alpha(\alpha+2\beta) \geq \beta^2$ holds since

$$\alpha(\alpha+2\beta) \geq \beta^2 \geq 3r(3t-\beta) \geq 3r(3t-\alpha-\beta).$$

Note also that the inequality $\alpha(\alpha+2\beta) \geq \beta^2$ holds, for instance, when $\alpha \geq \beta$ and more generally when the premium content packages of upstream firms are not strong complements. By the lemmas above, we can make the convention to present the subsequent analysis in a concise manner that (i) $\hat{\lambda} = 1$ when $\alpha(\alpha+2\beta) - 3r(3t-\alpha-\beta) \geq 0$, and (ii) $\bar{\lambda} = 1$ when $\beta^2 - 3r(3t-\beta) \geq 0$. Observe that we have $\tilde{\lambda} < \bar{\lambda} \leq \hat{\lambda}$ for all possible values of the parameters by this convention, together with the assumption just made.

We can obtain the following equilibrium characterization: We relegate the detailed analysis to Appendix A.

**Proposition 1.** *Assume that $\alpha(\alpha+2\beta) - 3r(3t-\alpha-\beta) \geq 0$ holds when $\beta^2 - 3r(3t-\beta) \geq 0$ holds.*

- *(i)* *When $\lambda < \tilde{\lambda}$: the unique Nash equilibrium (in fact, the dominant strategy equilibrium) is (E(o), E(o)). That is, the content providers offer the exclusive contract to different platforms.*
- *(ii)* *When $\tilde{\lambda} < \lambda < \bar{\lambda}$: the unique Nash equilibrium (in fact, the dominant strategy equilibrium) is (E(s), E(s)). That is, the content providers offer the exclusive contract to the same platform.*
- *(iii)* *When $\bar{\lambda} < \lambda < \hat{\lambda}$: there are two Nash equilibria, (E(s), E(s)) and (N,N).*[15]
- *(iv)* *When $\hat{\lambda} < \lambda$: the unique Nash equilibrium (in fact, the dominant strategy*

[15] There also exists a unique mixed strategy Nash equilibrium, in which each content provider chooses E with probability $\left(2\left(\beta^2 - 3r(3t-\beta)\right)\lambda + 3r(3t-\beta)\right)/\left((2\beta^2 - 2\alpha^2 - 4\alpha\beta - 6r\alpha)\lambda + 3r\alpha\right)$ and chooses N with the remaining probability. Throughout the paper, we mainly focus on the pure strategy Nash equilibria.

*equilibrium) is (N, N).*

*If, on the other hand,* $\beta^2 - 3r(3t - \beta) \geq 0$ *but* $\alpha(\alpha + 2\beta) - 3r(3t - \alpha - \beta) < 0$, *one content provider offers the exclusive contract while the other content provider offers the non-exclusive contract in equilibrium.*

Given the value of the parameter $\lambda$ representing the relative bargaining power of the upstream firm, we observe that the possibility of exclusive contracts rises when the cut-off values $\hat{\lambda}$ and $\bar{\lambda}$ are bigger, and we recall that (i) $\hat{\lambda}$ gets bigger when $\alpha$ increases, $\beta$ increases, $t$ decreases, and $r$ decreases, and (ii) $\bar{\lambda}$ gets bigger when $\beta$ increases, $t$ decreases, and $r$ decreases. Therefore,

**Theorem 1.** *The possibility of exclusive contracts rises when the value of the premium content increases (*$\alpha$ *and* $\beta$ *increase), the degree of horizontal differentiation in the downstream market decreases (*$t$ *decreases), and the importance of advertising revenue decreases (*$r$ *decreases).*

The platforms have a strong incentive to exclusively secure the premium content and gain a competitive advantage when the value of the premium content is high and/or the downstream competition is intense. Hence, the content providers can extract more surplus from the platforms when offering exclusive contracts. The opportunity cost of exclusive contracts for the content providers is small when the advertising revenue is low.[16]

Observe that, when $r = 0$, we have $\tilde{\lambda} = 0$ as well as $\beta^2 - 3r(3t - \beta) \geq 0$ and $\alpha(\alpha + 2\beta) - 3r(3t - \alpha - \beta) \geq 0$ hold and thus $\hat{\lambda} = \bar{\lambda} = 1$. This fits into case (ii) of Proposition 1 and we have the following result.

**Corollary 1.** *If there is no advertising revenue (*$r = 0$*), then, regardless of the relative bargaining power of the upstream firm, i.e., for all* $\lambda \in [0,1]$, *it is a dominant strategy for the content providers to offer the exclusive contract to the same platform.*

Therefore, the general analysis in this subsection encompasses the preliminary analysis above. We also have the following result by the fact that $\tilde{\lambda} < 1/2$, $\hat{\lambda} > 1/2$, and $\bar{\lambda} > 1/2$.

[16] We note that this comparative static result also holds true in Stennek's (2014) model with one monopolistic content provider.

**Corollary 2.** *If the upstream firm and the downstream firm hold equal bargaining power (*$\lambda \in 1/2$*), then, regardless of the magnitude of the advertising revenue, it is a dominant strategy for the content providers to offer the exclusive contract to the same platform.*

Stennek (2014) has shown, in a model with a monopolistic content provider and under the assumption that the upstream firm and the downstream firm hold equal bargaining power ($\lambda = 1/2$), that either exclusive or non-exclusive contracts may prevail depending on the values of the parameters. In contrast, we show in this model of competing content providers that exclusive contracts always prevail when $\lambda = 1/2$. Therefore, competition in the upstream market raises the possibility of exclusive contracts. The reason can be explained as follows. Due to upstream competition rather than upstream monopoly, the downstream firms' share of total industry profit gets larger and hence the downstream competition for exclusively securing the premium content intensifies. Consequently, the upstream firms can earn higher profits by offering exclusive contracts.

Regarding whether the content providers offer the exclusive contract to the same platform or to different platforms, we have the following result by the fact that $\tilde{\lambda}$ gets bigger when $\alpha$ decreases, $\beta$ decreases, and $r$ increases.

**Theorem 2.** *The possibility that the content providers offer the exclusive contract to the same platform rather than to different platforms rises when the value of the premium content increases (*$\alpha$ *and* $\beta$ *increase) and the importance of advertising revenue decreases (*$r$ *decreases).*

Finally, we observe that the parameter $\lambda$ also affects the contractual form.

**Theorem 3**. *The possibility of exclusive contracts rises when the relative bargaining power of the upstream firm gets weaker. Moreover, the content providers offer the exclusive contract to different platforms when the relative bargaining power is sufficiently weak (specifically when* $\lambda < \tilde{\lambda}$*).*

The reason for this result is similar to the one given just above. When the relative bargaining power of the downstream firms gets stronger, their industry profit gets larger and hence the downstream competition for exclusively securing the premium content intensifies. Consequently, the upstream firms can earn higher profits by offering exclusive contracts. When the relative bargaining power of the downstream firms is sufficiently strong, no platform gives in and each of them secures the premium content of one upstream firm.

It might appear that the general analysis is in contradiction to the earlier preliminary analysis in the following sense. Theorem 3 implies that the possibility of non-exclusive contracts rises when the relative bargaining power of the upstream firm gets stronger. In particular, the content providers offer the non-exclusive contract to the platforms when $\hat{\lambda} < \lambda$, as shown in case (iv) of Proposition 1. On the other hand, the preliminary analysis establishes that the content providers offer the exclusive contract when $\lambda = 1$. Observe however that this is not a contraction as Corollary 1 and the accompanying discussion demonstrates: We have $\tilde{\lambda} = 0$ and $\hat{\lambda} = \bar{\lambda} = 1$ when $r = 0$, thus the preliminary analysis fits into case (ii) of Proposition 1 and the content providers offer the exclusive contract to the same platform.

## 3. Two vertically integrated firms

In this section, we consider the situation where upstream firm A and downstream firm 1 are vertically integrated as well as upstream firm B and downstream firm 2 are vertically integrated. Hence, there are two vertically integrated firms, denoted by firm A1 and firm B2. Firm A1 consists of upstream sector A that owns the premium content and downstream sector 1 that competes for subscribers, and similarly for firm B2. Each firm's strategy is either to keep its own premium content exclusively or to make it available to the other firm. The former strategy is denoted by E and the latter strategy is denoted by N.

There are four possible strategy profiles, (E,E), (E,N), (N,E), and (N,N). We describe the resulting outcome for each of these strategy profiles.[17]

(1) When the strategy profile is (E,E):

The resulting downstream outcome is given as

$$v_1 = v_2 = v + \alpha; p_1 = p_2 = t; q_1 = q_2 = \frac{1}{2}; \pi_1 = \pi_2 = \frac{t}{2} ,$$

where the subscript $i = 1,2$ now represents the downstream sector. Since there is no trade of content between the firms, i.e., each firm keeps its own premium content exclusively, the resulting profits of the integrated firms are

$$\pi_{A1} = \pi_{B2} = \frac{t}{2} + \frac{r}{2} .$$

[17] We note that, unlike the previous section, we perform the analysis focusing on each strategy profile. We also note that we could alternatively adopt the current approach in the previous section and get the same results.

(2) When the strategy profile is (N,E):

The resulting downstream outcome is given as

$$v_1 = v + \alpha, v_2 = v + \alpha + \beta; p_1 = t - \frac{\beta}{3}, p_2 = t + \frac{\beta}{3};$$
$$q_1 = \frac{1}{2} - \frac{\beta}{6t}, q_2 = \frac{1}{2} + \frac{\beta}{6t}; \pi_1 = \frac{1}{2t}\left(t - \frac{\beta}{3}\right)^2, \pi_2 = \frac{1}{2t}\left(t + \frac{\beta}{3}\right)^2.$$

As for the upstream bargaining for content, with the disagreement outcome of A1's not providing its premium content, the gain of B2 is

$$\frac{1}{2t}\left(t + \frac{\beta}{3}\right)^2 - \frac{t}{2} + \left(\frac{1}{2} + \frac{\beta}{6t} - \frac{1}{2}\right) r$$

and the gain of A1 is

$$\frac{1}{2t}\left(t - \frac{\beta}{3}\right)^2 - \frac{t}{2} + \left(1 - \frac{1}{2}\right) r.$$

Hence,

$$l = \lambda \left(\frac{\beta}{3} + \frac{\beta^2}{18t} + \frac{\beta r}{6t}\right) - (1 - \lambda)\left(-\frac{\beta}{3} + \frac{\beta^2}{18t} + \frac{r}{2}\right),$$

$$\pi_{A1} = \frac{1}{2t}\left(t - \frac{\beta}{3}\right)^2 + r + l = \frac{t}{2} + \frac{r}{2} + \lambda\left(\frac{\beta^2}{9t} + \frac{\beta r}{6t} + \frac{r}{2}\right), \text{ and}$$

$$\pi_{B2} = \frac{1}{2t}\left(t + \frac{\beta}{3}\right)^2 + \left(\frac{1}{2} + \frac{\beta}{6t}\right) r - l = \frac{t}{2} + \frac{\beta^2}{9t} + \left(1 + \frac{\beta}{6t}\right) r - \lambda\left(\frac{\beta^2}{9t} + \frac{\beta r}{6t} + \frac{r}{2}\right).$$

(3) When the strategy profile is (E,N):

The outcome is the mirror image of that of (N,E).

(4) When the strategy profile is (N,N):

The resulting downstream outcome is given as

$$v_1 = v_2 = v + \alpha + \beta; p_1 = p_2 = t; q_1 = q_2 = \frac{1}{2}; \pi_1 = \pi_2 = \frac{t}{2}.$$

Since the lump-sum fees for the premium content between the firms cancel out,

$$\pi_{A1} = \pi_{B2} = \frac{t}{2} + r.$$

Summarizing, the game matrix is given as

| | E | N |
|---|---|---|
| E | $\frac{t}{2}+\frac{r}{2}$, $\frac{t}{2}+\frac{r}{2}$ | $\frac{t}{2}+\frac{\beta^2}{9t}+\left(1+\frac{\beta}{6t}\right)r-\lambda\left(\frac{\beta^2}{9t}+\frac{\beta r}{6t}+\frac{r}{2}\right)$, $\frac{t}{2}+\frac{r}{2}+\lambda\left(\frac{\beta^2}{9t}+\frac{\beta r}{6t}+\frac{r}{2}\right)$ |
| N | $\frac{t}{2}+\frac{r}{2}+\lambda\left(\frac{\beta^2}{9t}+\frac{\beta r}{6t}+\frac{r}{2}\right)$, $\frac{t}{2}+\frac{\beta^2}{9t}+\left(1+\frac{\beta}{6t}\right)r-\lambda\left(\frac{\beta^2}{9t}+\frac{\beta r}{6t}+\frac{r}{2}\right)$ | $\frac{t}{2}+r$, $\frac{t}{2}+r$ |

It is straightforward to obtain the following result.[18]

**Proposition 2.** *There are two Nash equilibria, (N,E) and (E,N), when the inequality*

$$\lambda < \frac{2\beta^2 + 3\beta r}{2\beta^2 + 3\beta r + 9rt}$$

*holds. On the other hand, strategy N is a dominant strategy when the reverse inequality holds.*

As for the comparative statics with respect to the parameters, we have:

**Theorem 4.** *The possibility of exclusive contracts rises when the value of the premium content increases ($\beta$ increases), the degree of horizontal differentiation in the downstream market decreases ($t$ decreases), the importance of advertising revenue decreases ($r$ decreases), and the relative bargaining power of the upstream firm gets weaker ($\lambda$ decreases).*

## 4. One vertically integrated firm

In this section, we consider the situation where upstream firm A and downstream firm 1 are vertically integrated whereas firm B and firm 2 are independent. Hence, there are three firms in the market: one vertically integrated firm A1, one independent upstream firm B, and one independent downstream firm 2. Firm A1's strategy is either to keep its own premium content exclusively or to make it available to firm 2. The former strategy is denoted by E and the latter strategy is denoted by N. Firm B's strategy is to offer the exclusive contract to firm 2, to offer the exclusive contract to firm A1, or to offer the non-exclusive contract to both firm 2 and firm A1. These strategies are denoted by $E_2$, $E_{A1}$, and N, respectively.

[18] If $\lambda = 0$, then (E,E) is also an equilibrium. We ignore this case.

There are six possible strategy profiles, (E,$E_2$), (E,$E_{A1}$), (E,N), (N,$E_2$), (N,$E_{A1}$), and (N,N). Since the analysis is rather mechanical, we relegate a general analysis to Appendix B and report in this section only the results for the case when there is no advertising revenue of the upstream firm, i.e., $r = 0$.[19] Then, the game matrix is given as follows. In addition, we exclude the case when $\lambda = 0$, i.e., we assume that $\lambda \in (0,1]$ to avoid trivialities: Note in particular that firm B's payoff is zero in all strategy profiles when $\lambda = 0$.

| | $E_2$ | $E_{A1}$ | N |
|---|---|---|---|
| E | $\frac{t}{2}$, $\frac{\lambda(\alpha+\beta)(6t-\alpha-\beta)}{18t}$ | $\frac{1}{2t}\left(t+\frac{\alpha+\beta}{3}\right)^2 - \frac{\lambda(\alpha+\beta)(6t+\alpha+\beta)}{18t}$, $\frac{\lambda(\alpha+\beta)(6t+\alpha+\beta)}{18t}$ | $\frac{1}{2t}\left(t+\frac{\beta}{3}\right)^2 - \lambda\left(\frac{\beta}{3}+\frac{\beta^2}{18t}\right)$, $\lambda\left(\frac{\alpha+\beta}{3}+\frac{\beta^2-\alpha^2-2\alpha\beta}{18t}\right)$ |
| N | $\frac{t}{2}+\lambda\frac{\beta^2}{9t}$, $\lambda\frac{2\beta}{3}$ | $\frac{1}{2t}\left(t+\frac{\alpha+\beta}{3}\right)^2 - \lambda\left(\frac{\alpha(\alpha+2\beta)}{9t}+\frac{2\beta}{3}\right)$, $\lambda\frac{2\beta}{3}$ | $\frac{1}{2t}\left(t+\frac{\beta}{3}\right)^2 - \lambda\left(\frac{\beta}{3}+\frac{\beta^2}{18t}\right)$, $2\lambda\left(\frac{\beta}{3}-\frac{\beta^2}{18t}\right)$ |

It is immediate to see that strategy N is firm A1's best response to firm B's strategy $E_2$, and both E and N are firm A1's best responses to firm B's strategy N. As for firm B's strategy $E_{A1}$, strategy E is a best response if

$$\alpha^2 + 2\alpha\beta - \beta^2 > 6t(\alpha - \beta),$$

holds and strategy N is a best response if the reverse inequality holds. Observe that this inequality holds when $\alpha = \beta$. Next, it is easy to see that strategy $E_{A1}$ is a best response to firm A1's strategy E, and both $E_2$ and $E_{A1}$ are best responses to firm A1's strategy N. Summarizing the discussion,

[19] When $r > 0$, there are too many cases to consider depending on the relative values of the parameters.

**Proposition 3.** *Assume that* $r = 0$.

*(i)* *When* $\alpha^2 + 2\alpha\beta - \beta^2 > 6t(\alpha - \beta)$*: there are two Nash equilibria in this game, the strategy profile (N,* $E_2$*) and the strategy profile (E,* $E_{A1}$*).*

*(ii)* *When* $\alpha^2 + 2\alpha\beta - \beta^2 < 6t(\alpha - \beta)$*: there are two Nash equilibria in this game, the strategy profile (N,* $E_2$*) and the strategy profile (N,* $E_{A1}$*).*

Unlike the cases of vertical separation and two vertical integrations, the comparative static results with respect to the parameters are not clear-cut. Observe from the inequality above that the possibility of exclusive contracts rises when $\alpha$ decreases but $\beta$ increases. It may rise or fall in $t$ depending on whether $\alpha > \beta$ or $\alpha < \beta$. Hence, we see that complicated strategic effects are at work when firms are asymmetric in vertical structure.

## 5. Comparison across vertical structures

In this section, we compare the vertical structures studied in Sections 2-4 and briefly examine firms' incentives to vertically integrate as well as the welfare implications. We assume that there is no advertising revenue, i.e., $\mathrm{r} = 0$, to facilitate the analysis.

Observe first from Proposition 1 and Corollary 1 in Section 2 that the only equilibrium for the case of vertically independent firms is (E(s), E(s)). Observe next from Proposition 2 in Section 3 that there are two equilibria, (N, E) and (E, N), for the case of two vertically integrated firms.[20] Finally for the case of one vertically integrated firm, Proposition 3 in Section 4 shows that there are (i) two equilibria, (N, $\mathrm{E}_2$) and (E, $\mathrm{E}_{\mathrm{A1}}$), if $\alpha^2 + 2\alpha\beta - \beta^2 > 6t(\alpha - \beta)$ holds, and (ii) two equilibria, (N, $\mathrm{E}_2$) and (N, $\mathrm{E}_{\mathrm{A1}}$), if the reverse inequality holds.

### 5-1. The merger involving the downstream firm with premium content

Let us first consider the vertical merger between an upstream firm and the downstream firm that secures the premium content of both upstream firms. Without loss of generality, consider the merger between firm A and firm 1.

[20] When $\lambda = 1$, we also have (N,N) as an equilibrium. We ignore this equilibrium since it is a fragile one occurring only at the boundary as well as the sum of firms' payoffs in this equilibrium is lower than that in the other equilibria.

We have:[21]

**Proposition 4.** *Firm A and firm 1 have an incentive to merge if either*

*(i)* *the expected equilibrium at one vertical integration is* (N, $E_2$) and

$$\lambda > \frac{(\alpha+\beta)(6t+\alpha+\beta)}{(\alpha+\beta)(6t+\alpha+\beta)+2\beta^2} ,$$

*or*

*(ii)* *the expected equilibrium at one vertical integration is* $(N,\ E_{A1})$, *which is indeed an equilibrium only when* $\alpha^2+2\alpha\beta-\beta^2<6t(\alpha-\beta)$.

We next consider the incentive for a counter-merger. That is, do firm B and firm 2 have an incentive to merge after the merger between firm A and firm 1? We have:

**Proposition 5.** *If the expected equilibrium at two vertical integrations is (E,N), then firm B and firm 2 have an incentive to counter-merge. If the expected equilibrium at two vertical integrations is (N,E), then firm B and firm 2 do not have an incentive to counter-merge for most parameter values. Anticipating the counter-merger, firm A and firm 1 no longer have an incentive to merge initially if the expected equilibrium at two vertical integrations is (E,N), whereas they still have an incentive to merge initially for most values of the parameters if the expected equilibrium at two vertical integrations is (N,E).*

In particular,

**Corollary 3.** *Assume that* $\alpha^2+2\alpha\beta-\beta^2>6t(\alpha-\beta)$. *With the possibility of merger between firm A and firm 1 and the counter-merger between firm B and firm 2, the only eventual market structures are (i) vertical separation and the equilibrium of (E(s), E(s)), and (ii) one vertical integration and the equilibrium of (N,* $E_2$*). Moreover, if* $\lambda=1/2$ *then the only eventual market structure is vertical separation and the equilibrium of (E(s), E(s)), i.e., there does not occur any actual merger.*

[21] To contain the length of the paper, we report only the main results in this section. A detailed analysis can be provided upon request.

## 5-2. The merger involving the downstream firm without premium content

Let us next consider the vertical merger between an upstream firm and the downstream firm that secures no premium content. Without loss of generality, consider the merger between firm B and firm 2.

**Proposition 6.** *Assume that* $\alpha^2 + 2\alpha\beta - \beta^2 > 6t(\alpha - \beta)$. *Then, with the possibility of merger between firm B and firm 2 and the counter-merger between firm A and firm 1,*

*(i)* *the only eventual market structure is vertical separation and the equilibrium of (E(s), E(s)), i.e., there does not occur any actual merger, when the relative bargaining power of the upstream firm is sufficiently strong, whereas*

*(ii)* *the most plausible market structure is two vertical integrations when the relative bargaining power of the upstream firm is sufficiently weak.*

From Propositions 5 and 6, we can get the following broad implications. If the bargaining power of the upstream firm is strong, then either no merger or the merger involving the downstream firm with premium content occurs. Hence, the resulting market structure is either the vertical separation in which both upstream firms offer the exclusive contract to the same downstream firm or one vertical integration in which the integrated firm provides its premium content to the unintegrated downstream firm. If the bargaining power of the upstream firm is weak, then the merger involving the firm with no premium content and the subsequent counter-merger occur. Hence, the resulting market structure is two vertical integrations.

In the merger involving the downstream firm with premium content, the *input supply effect* dominates the *downstream competition effect*, i.e., the merger occurs mainly to profit from the additional sale of the premium content. On the other hand, in the merger involving the downstream firm with no premium content, the reverse holds, i.e., the merger occurs mainly to secure the premium content and gain a competitive advantage in the downstream market. In any case, vertical integration leads to less exclusive content provision than vertical separation. We note that this is in nice contrast with D'Annunzio (2017), which has shown that the content provider always offers the exclusive contract whether it is vertically integrated or not (i.e., regardless of the vertical structure) in a model with one monopolistic content provider who makes a take-it-or-leave-it offer.

### 5-3. Welfare analysis

Let us compare the market performance in various vertical structure. In the unique equilibrium (E(s), E(s)) of vertical separation, the consumer surplus is

$$cs^{(E(s),E(s))} = \int_0^{q_1}(v+\alpha+\beta-tx-p_1)\,dx + \int_{q_1}^{1}(v-t(1-x)-p_2)\,dx$$

$$= \frac{(\alpha+\beta)^2+18(\alpha+\beta)t+9t(4v-5t)}{36t}$$

and the social welfare is

$$sw^{(E(s),E(s))} = cs^{(E(s),E(s))} + \pi_1 + \pi_2 = \frac{5(\alpha+\beta)^2+18(\alpha+\beta)t+9t(4v-t)}{36t}.$$

As for the equilibrium (N,$E_2$) of one vertical integration, the consumer surplus is

$$cs^{(N,E_2)} = \int_0^{q_1}(v+\alpha-tx-p_1)\,dx + \int_{q_1}^{1}(v+\alpha+\beta-t(1-x)-p_2)\,dx$$

$$= \frac{\beta^2+18(2\alpha+\beta)t+9t(4v-5t)}{36t}$$

and the social welfare is

$$sw^{(N,E_2)} = cs^{(N,E_2)} + \pi_1 + \pi_2 = \frac{5\beta^2+18(2\alpha+\beta)t+9t(4v-t)}{36t}.$$

It is straightforward to see that the consumer surplus and the social welfare at the equilibrium (E,$E_{A1}$) and (N,$E_{A1}$) of one vertical integration, respectively, are the same as those at (E(s),E(s)) and (N,$E_2$). Moreover, the consumer surplus and the social welfare at the equilibria (N,E) and (E,N) of two vertical integration are also the same as those at (N,$E_2$).

We have

$$cs^{(N,E_2)} - cs^{(E(s),E(s))} = \frac{\alpha(18t-\alpha-2\beta)}{36t} > 0,$$

but

$$sw^{(N,E_2)} - sw^{(E(s),E(s))} = \frac{\alpha(18t-5\alpha-10\beta)}{36t}.$$

Hence, the consumer surplus is higher at (N,$E_2$) than at (E(s),E(s)). For the social welfare, the comparison is ambiguous between (N,$E_2$) and (E(s),E(s)). Observe that $sw^{(N,E_2)} < sw^{(E(s),E(s))}$ when $t$ is relatively low compared to $\alpha$ and $\beta$. The reason is that, when the horizontal competition is intense due to low transport cost, (E(s),E(s)) is a better market structure to eschew the vertical competition and generate a higher industry profit.

## 6. Conclusion

With a multilateral vertical contracting model of media markets, we have examined the upstream competition and the contractual arrangements in content provision. We have analyzed the trade of content by the Nash bargaining solution and the downstream competition by the Hotelling location model. We have characterized the equilibrium outcomes and the contractual arrangements for various vertical structures, i.e., for vertical separation, partial vertical integration, and full vertical integration. We have also briefly examined firms' incentives to vertically integrate as well as the welfare implications.

We have shown that the possibility of exclusive contracts rises when the value of the premium content increases, the degree of horizontal differentiation in the downstream market decreases, the importance of advertising revenue decreases, and the relative bargaining power of the upstream firm gets weaker both under vertical separation and full vertical integration. We have also shown that, under vertical separation, the possibility that the content providers offer exclusive contracts to the same platform rather than to different platforms rises when the value of the premium content increases and the importance of upstream advertising revenue decreases, but the reverse holds true when the relative bargaining power of the upstream firm is sufficiently weak. While unambiguous comparative static results are impossible to obtain for partial vertical integration, we have found that there are only three possible contractual arrangements.

We have shown that (i) if the bargaining power of the upstream firm is strong then there occurs no merger or the merger involving the downstream firm with premium content whereas (ii) if the bargaining power of the upstream firm is weak then there occur the merger involving the downstream firm with no premium content and the subsequent counter-merger, resulting in full vertical integration. Since firms have an incentive to merge either to profit from the additional sale of the premium content or to compete more effectively in the downstream market in our model, merger generally induces less exclusive content provision and enhances welfare.[22]

We have obtained these results with a highly stylized model, mainly to concentrate on upstream competition and the effect of relative bargaining power. First of all, we have assumed that consumers subscribe to one and only one platform, that is, they single-home. Since media platforms such as multichannel video programming distributors (MVPDs) and online video distributors (OVDs) charge considerable subscription fees to consumers, it certainly is a costly decision to subscribe to more than one platform. This is in nice contrast to many other platforms including social networking services and online marketplaces to which consumers can subscribe for free. Nevertheless, media consumers these days may subscribe to more than one platform. For instance, one may be a customer of both Netflix and

---

[22] We hasten to add, however, that this conclusion should be interpreted with care. In particular, the current model does consider the foreclosure effect nor the investment incentives.

Amazon Prime Video. Hence, it is worthwhile to extend the model to the case when consumers multi-home.

Secondly, we have assumed that the upstream content providers earn advertising revenues but the downstream platforms do not. This assumption may be reasonable when the upstream firms are sports content providers and the downstream firms are online video distributors. In addition, product placement in movies and drama series instead of commercial breaks is popular in media markets. On the other hand, the downstream firms may also earn advertising revenues when they are broadcast stations or multichannel video programming distributors (MVPDs). We admit that this assumption is made mainly to facilitate the analysis, and it is a future research agenda to extend the model to the case when the downstream firms earn advertising revenues.

## Appendix A: The equilibrium characterization of vertically independent firms

[Case 1] When $\lambda < \tilde{\lambda}$ holds:

The game matrix is given as follows.

| | E(o) | N |
|---|---|---|
| E(o) | $\frac{\lambda(\alpha+\beta)(6t-3r-\alpha-\beta)}{18t}+\left(\frac{1}{2}+\frac{\alpha+\beta}{6t}\right)r$, $\frac{\lambda(\alpha+\beta)(6t-3r-\alpha-\beta)}{18t}+\left(\frac{1}{2}+\frac{\alpha+\beta}{6t}\right)r$ | $\lambda\frac{2\beta}{3}+\left(\frac{1}{2}+\frac{\beta}{6t}\right)r$, $\frac{\lambda(6(\alpha+\beta)t+\beta^2-\alpha^2-2\alpha\beta)}{18t}+\frac{(\lambda(6t-\alpha-\beta)+(\alpha+\beta))r}{6t}$ |
| N | $\frac{\lambda(6(\alpha+\beta)t+\beta^2-\alpha^2-2\alpha\beta)}{18t}+\frac{(\lambda(6t-\alpha-\beta)+(\alpha+\beta))r}{6t}$, $\lambda\frac{2\beta}{3}+\left(\frac{1}{2}+\frac{\beta}{6t}\right)r$ | $\lambda\left(\frac{2\beta}{3}-\frac{\beta^2}{9t}+\left(1-\frac{\beta}{3t}\right)r\right)+\frac{\beta r}{3t}$, $\lambda\left(\frac{2\beta}{3}-\frac{\beta^2}{9t}+\left(1-\frac{\beta}{3t}\right)r\right)+\frac{\beta r}{3t}$ |

By Lemmas 1 and 2, the unique Nash equilibrium (in fact, the dominant strategy equilibrium) is (E(o), E(o)). That is, the content providers offer the exclusive contract to different platforms. Diagrammatically,

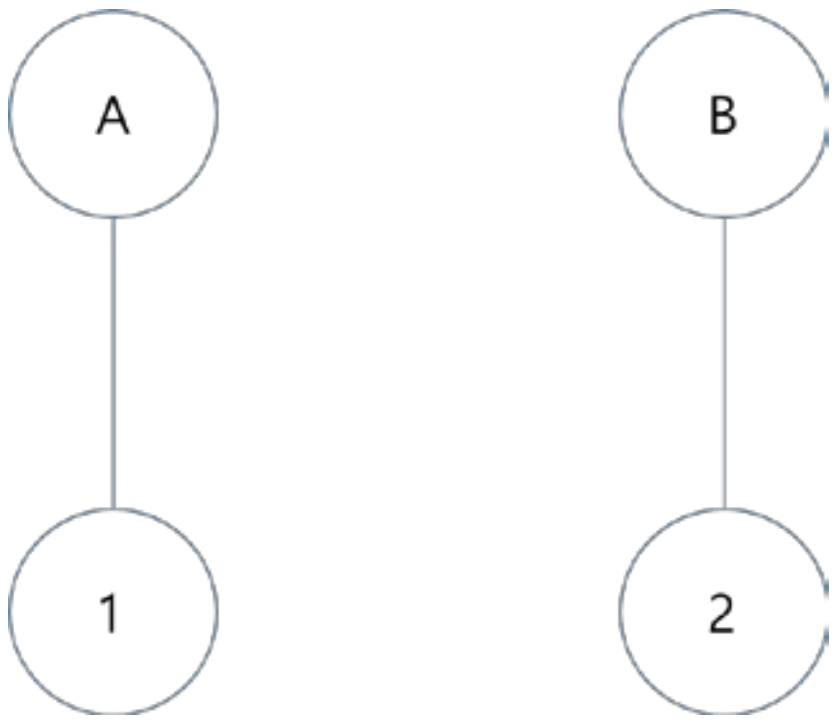

[Case 2] When $\tilde{\lambda} < \lambda < \bar{\lambda}$ holds:

The game matrix is given as follows.

| | E(s) | N |
|---|---|---|
| E(s) | $\frac{\lambda(\alpha+\beta)(6t+3r+\alpha+\beta)}{18t}+\frac{r}{2}$ , $\frac{\lambda(\alpha+\beta)(6t+3r+\alpha+\beta)}{18t}+\frac{r}{2}$ | $\lambda\frac{2\beta}{3}+\left(\frac{1}{2}+\frac{\beta}{6t}\right)r,$ $\frac{\lambda(6(\alpha+\beta)t+\beta^2-\alpha^2-2\alpha\beta)}{18t}+\frac{(\lambda(6t-\alpha-\beta)+(\alpha+\beta))r}{6t}$ |
| N | $\frac{\lambda(6(\alpha+\beta)t+\beta^2-\alpha^2-2\alpha\beta)}{18t}+\frac{(\lambda(6t-\alpha-\beta)+(\alpha+\beta))r}{6t}$, $\lambda\frac{2\beta}{3}+\left(\frac{1}{2}+\frac{\beta}{6t}\right)r$ | $\lambda\left(\frac{2\beta}{3}-\frac{\beta^2}{9t}+\left(1-\frac{\beta}{3t}\right)r\right)+\frac{\beta r}{3t}$ , $\lambda\left(\frac{2\beta}{3}-\frac{\beta^2}{9t}+\left(1-\frac{\beta}{3t}\right)r\right)+\frac{\beta r}{3t}$ |

By Lemmas 1 and 2, the unique Nash equilibrium (in fact, the dominant strategy equilibrium) is (E(s), E(s)). That is, the content providers offer the exclusive contract to the same platform. Diagrammatically,

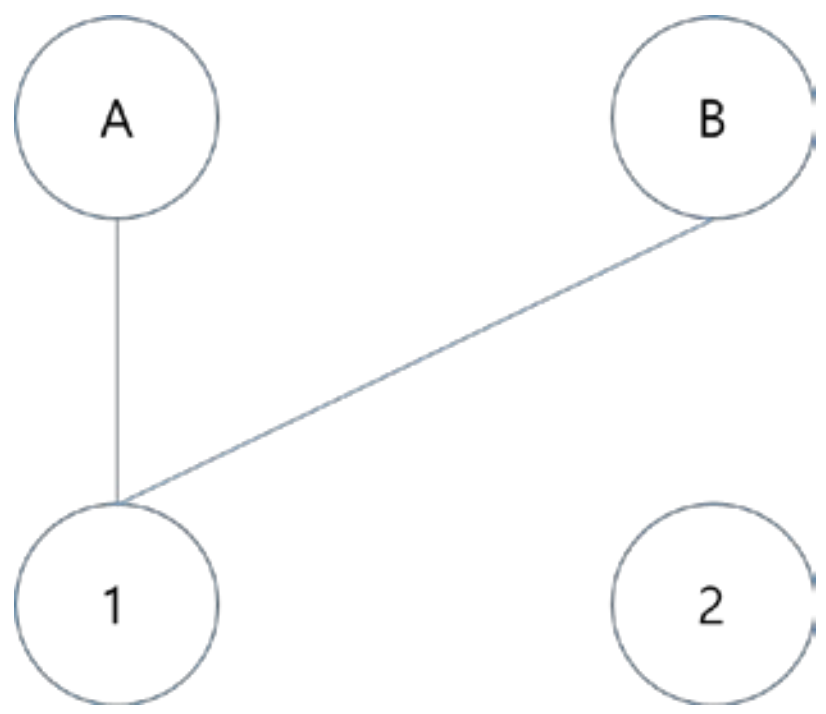

[Case 3] When $\bar{\lambda} < \lambda < \hat{\lambda}$ holds:

The game matrix is the same as that in Case 2. By Lemmas 1 and 2, there are two Nash equilibria, (E(s), E(s)) and (N,N).[23]

[23] There also exists a unique mixed strategy Nash equilibrium, in which each content provider chooses E with

[Case 4] When $\hat{\lambda} < \lambda$ holds:

The game matrix is the same as that in Case 2. By Lemmas 1 and 2, the unique Nash equilibrium (in fact, the dominant strategy equilibrium) is (N, N). That is, the content providers offer the non-exclusive contract to the platforms. Diagrammatically,

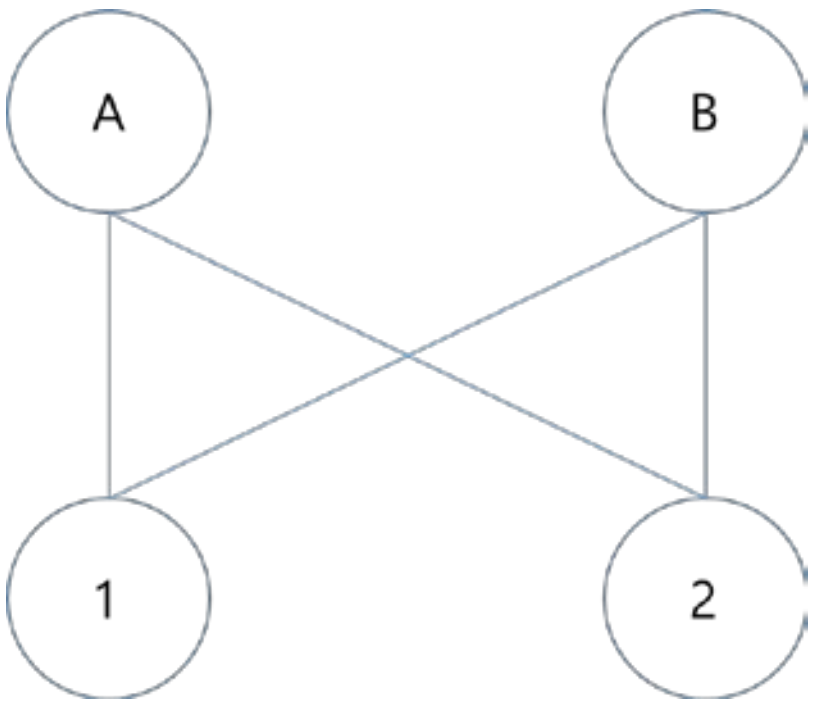

Let us also discuss the case when

$$\beta^2 - 3r(3t - \beta) \geq 0 \text{ but } \alpha(\alpha + 2\beta) - 3r(3t - \alpha - \beta) < 0,$$

which corresponds to the case when $\hat{\lambda} < \bar{\lambda}$ holds in our convention. When $\hat{\lambda} < \lambda < \bar{\lambda}$, or more precisely, when $\alpha(\alpha + 2\beta) - 3r(3t - \alpha - \beta) < 0$, $\hat{\lambda} < \lambda$, and $\beta^2 - 3r(3t - \beta) \geq 0$ hold, by Lemmas 1 and 2, there are two Nash equilibria (E,N) and (N,E) of the game matrix given in Case 2. Hence, one content provider offers the exclusive contract while the other content provider offers the non-exclusive contract. The equilibrium (E,N) is depicted below.[24]

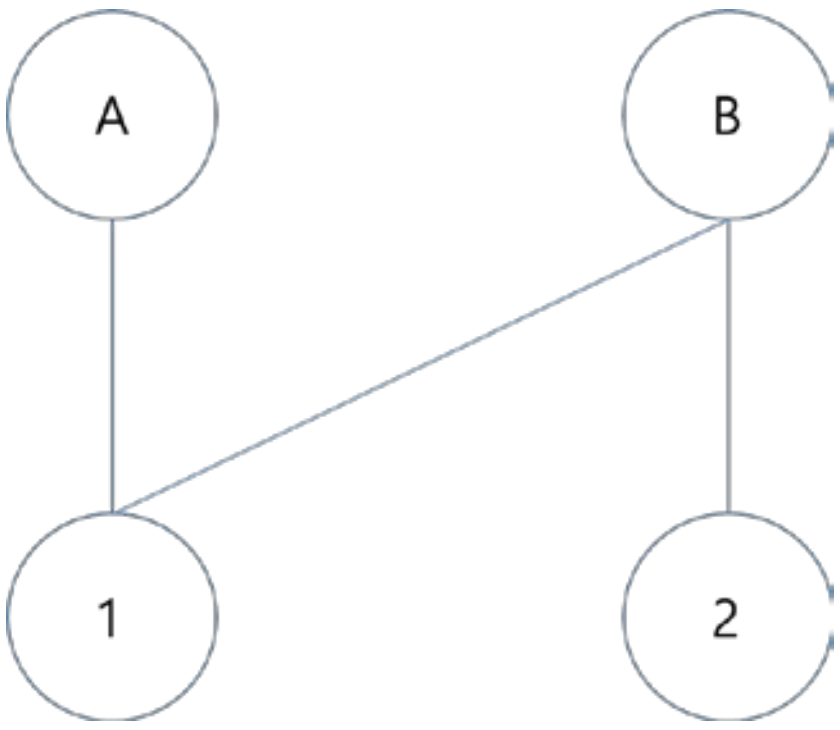

---

probability $\left(2(\beta^2 - 3r(3t - \beta))\lambda + 3r(3t - \beta)\right)/\left((2\beta^2 - 2\alpha^2 - 4\alpha\beta - 6r\alpha)\lambda + 3r\alpha\right)$ and chooses N with the remaining probability. Throughout the paper, we mainly focus on the pure strategy Nash equilibria.

[24] The equilibrium (N,E) can be symmetrically depicted.

## Appendix B: The analysis of one vertical integration

(1) When the strategy profile is (E,$E_2$):

The resulting downstream outcome is given as

$$v_1 = v_2 = v + \alpha; p_1 = p_2 = t; q_1 = q_2 = \frac{1}{2}; \pi_1 = \pi_2 = \frac{t}{2}\ .$$

Firm A1's profit is

$$\pi_{A1} = \frac{t}{2}\ .$$

In the negotiation between B and 2, with the disagreement outcome of B's providing its premium content exclusively to A1, the gain of 2 is

$$\frac{t}{2} - \frac{1}{2t}\left(t - \frac{\alpha+\beta}{3}\right)^2$$

and the gain of B is

$$\frac{r}{2} - \left(\frac{1}{2} + \frac{\alpha+\beta}{6t}\right) r.$$

Hence,

$$l = \lambda\left(\frac{\alpha+\beta}{3} - \frac{(\alpha+\beta)^2}{18t}\right) - (1-\lambda)\left(-\frac{\alpha+\beta}{6t} r\right),$$

$$\pi_B = \frac{r}{2} + l = \frac{\lambda(\alpha+\beta)(6t-3r-\alpha-\beta)}{18t} + \left(\frac{1}{2} + \frac{\alpha+\beta}{6t}\right) r.$$

(2) When the strategy profile is (E,$E_{A1}$):

The resulting downstream outcome is given as

$$v_1 = v + \alpha + \beta, v_2 = v; p_1 = t + \frac{\alpha+\beta}{3}, p_2 = t - \frac{\alpha+\beta}{3};$$

$$q_1 = \frac{1}{2} + \frac{\alpha+\beta}{6t}, q_2 = \frac{1}{2} - \frac{\alpha+\beta}{6t}; \pi_1 = \frac{1}{2t}\left(t + \frac{\alpha+\beta}{3}\right)^2, \pi_2 = \frac{1}{2t}\left(t - \frac{\alpha+\beta}{3}\right)^2.$$

In the negotiation between B and A1, with the disagreement outcome of B's providing its premium content exclusively to 2, the gain of A1 is

$$\frac{1}{2t}\left(t + \frac{\alpha+\beta}{3}\right)^2 - \frac{t}{2} + \left(\frac{1}{2} + \frac{\alpha+\beta}{6t} - \frac{1}{2}\right) r$$

and the gain of B is

$$\left(\frac{1}{2}+\frac{\alpha+\beta}{6t}-\frac{1}{2}\right)r=\frac{\alpha+\beta}{6t}r.$$

Hence,

$$\begin{aligned}
l &= \lambda\left(\frac{\alpha+\beta}{3}+\frac{(\alpha+\beta)^2}{18t}+\frac{\alpha+\beta}{6t}r\right)-(1-\lambda)\frac{\alpha+\beta}{6t}r,\\
\pi_{A1} &= \frac{1}{2t}\left(t+\frac{\alpha+\beta}{3}\right)^2+\left(\frac{1}{2}+\frac{\alpha+\beta}{6t}\right)r-l\\
&= \frac{1}{2t}\left(t+\frac{\alpha+\beta}{3}\right)^2+\left(\frac{1}{2}+\frac{\alpha+\beta}{3t}\right)r-\lambda\left(\frac{\alpha+\beta}{3}+\frac{(\alpha+\beta)^2}{18t}+\frac{\alpha+\beta}{3t}r\right),\\
\pi_B &= \left(\frac{1}{2}+\frac{\alpha+\beta}{6t}\right)r+l=\frac{\lambda(\alpha+\beta)(6t+6r+\alpha+\beta)}{18t}+\frac{r}{2}\ .
\end{aligned}$$

(3) When the strategy profile is (E,N):

The resulting downstream outcome is given as

$$v_1=v+\alpha+\beta, v_2=v+\alpha; p_1=t+\frac{\beta}{3}, p_2=t-\frac{\beta}{3};$$
$$q_1=\frac{1}{2}+\frac{\beta}{6t}, q_2=\frac{1}{2}-\frac{\beta}{6t}; \pi_1=\frac{1}{2t}\left(t+\frac{\beta}{3}\right)^2, \pi_2=\frac{1}{2t}\left(t-\frac{\beta}{3}\right)^2.$$

In the negotiation between B and A1, with the disagreement outcome of B’s providing its premium content only to 2, the gain of A1 is

$$\frac{1}{2t}\left(t+\frac{\beta}{3}\right)^2-\frac{t}{2}+\left(\frac{1}{2}+\frac{\beta}{6t}-\frac{1}{2}\right)r$$

and the gain of B is

$$\left(1-\frac{1}{2}\right)r=\frac{r}{2}\ .$$

Hence,

$$l_{B,A1}=\lambda\left(\frac{\beta}{3}+\frac{\beta^2}{18t}+\frac{\beta r}{6t}\right)-(1-\lambda)\frac{r}{2}\ .$$

In the negotiation between B and 2, with the disagreement outcome of B’s providing its premium content only to A1, the gain of 2 is

$$\frac{1}{2t}\left(t-\frac{\beta}{3}\right)^2-\frac{1}{2t}\left(t-\frac{\alpha+\beta}{3}\right)^2$$

and the gain of B is

$$\left(1-\frac{1}{2}-\frac{\alpha+\beta}{6t}\right)r=\left(\frac{1}{2}-\frac{\alpha+\beta}{6t}\right)r.$$

Hence,

$$l_{B,2}=\lambda\left(\frac{\alpha}{3}-\frac{\alpha^2+2\alpha\beta}{18t}\right)-(1-\lambda)\left(\frac{1}{2}-\frac{\alpha+\beta}{6t}\right)r.$$

We have

$$\begin{aligned}\pi_{A1}&=\frac{1}{2t}\left(t+\frac{\beta}{3}\right)^2+\left(\frac{1}{2}+\frac{\beta}{6t}\right)r-l_{B,A1}\\&=\frac{1}{2t}\left(t+\frac{\beta}{3}\right)^2+\left(1+\frac{\beta}{6t}\right)r-\lambda\left(\frac{\beta}{3}+\frac{\beta^2}{18t}+\left(\frac{1}{2}+\frac{\beta}{6t}\right)r\right),\\\pi_B&=r+l_{B,A1}+l_{B,2}\\&=\frac{\alpha+\beta}{6t}r+\lambda\left(\frac{\alpha+\beta}{3}+\frac{\beta^2-\alpha^2-2\alpha\beta}{18t}+\left(1-\frac{\alpha}{6t}\right)r\right).\end{aligned}$$

(4) When the strategy profile is (N,$E_2$):

The resulting downstream outcome is given as

$$v_1=v+\alpha,v_2=v+\alpha+\beta;p_1=t-\frac{\beta}{3},p_2=t+\frac{\beta}{3};$$
$$q_1=\frac{1}{2}-\frac{\beta}{6t},q_2=\frac{1}{2}+\frac{\beta}{6t};\pi_1=\frac{1}{2t}\left(t-\frac{\beta}{3}\right)^2,\pi_2=\frac{1}{2t}\left(t+\frac{\beta}{3}\right)^2.$$

In the negotiation between A1 and 2, with the disagreement outcome of A1's not providing its premium content to 2, the gain of 2 is

$$\frac{1}{2t}\left(t+\frac{\beta}{3}\right)^2-\frac{t}{2}$$

and the gain of A1 is

$$\frac{1}{2t}\left(t-\frac{\beta}{3}\right)^2-\frac{t}{2}+\left(1-\frac{1}{2}\right)r.$$

Hence,

$$l_{A1,2}=\lambda\left(\frac{\beta}{3}+\frac{\beta^2}{18t}\right)-(1-\lambda)\left(-\frac{\beta}{3}+\frac{\beta^2}{18t}+\frac{r}{2}\right).$$

In the negotiation between B and 2, with the disagreement outcome of B's providing its premium content exclusively to A1, the gain of 2 is

$$\frac{1}{2t}\left(t+\frac{\beta}{3}\right)^2-\frac{1}{2t}\left(t-\frac{\beta}{3}\right)^2$$

and the gain of B is

$$\left(\frac{1}{2}+\frac{\beta}{6t}-\left(\frac{1}{2}+\frac{\beta}{6t}\right)\right)r=0.$$

Hence,

$$l_{B,2}=\lambda\frac{2\beta}{3}\ .$$

We have

$$\pi_{A1}=\frac{1}{2t}\left(t-\frac{\beta}{3}\right)^2+r+l_{A1,2}=\frac{t}{2}+\frac{r}{2}+\lambda\left(\frac{\beta^2}{9t}+\frac{r}{2}\right),$$

$$\pi_B=\lambda\frac{2\beta}{3}\ .$$

(5) When the strategy profile is (N,$E_{A1}$):

The resulting downstream outcome is given as

$$v_1=v+\alpha+\beta, v_2=v+\alpha; p_1=t+\frac{\beta}{3}, p_2=t-\frac{\beta}{3};$$
$$q_1=\frac{1}{2}+\frac{\beta}{6t}, q_2=\frac{1}{2}-\frac{\beta}{6t}; \pi_1=\frac{1}{2t}\left(t+\frac{\beta}{3}\right)^2, \pi_2=\frac{1}{2t}\left(t-\frac{\beta}{3}\right)^2.$$

In the negotiation between A1 and 2, with the disagreement outcome of A1's not providing its premium content to 2, the gain of 2 is

$$\frac{1}{2t}\left(t-\frac{\beta}{3}\right)^2-\frac{1}{2t}\left(t-\frac{\alpha+\beta}{3}\right)^2$$

and the gain of A1 is

$$\frac{1}{2t}\left(t+\frac{\beta}{3}\right)^2-\frac{1}{2t}\left(t+\frac{\alpha+\beta}{3}\right)^2+\left(1-\frac{1}{2}-\frac{\alpha+\beta}{6t}\right)r.$$

Hence,

$$l_{A1,2}=\lambda\left(\frac{\alpha}{3}-\frac{\alpha^2+2\alpha\beta}{18t}\right)-(1-\lambda)\left(-\frac{\alpha}{3}-\frac{\alpha^2+2\alpha\beta}{18t}+\left(\frac{1}{2}-\frac{\alpha+\beta}{6t}\right)r\right).$$

In the negotiation between B and A1, with the disagreement outcome of B's providing its premium content exclusively to 2, the gain of A1 is

$$\frac{1}{2t}\left(t+\frac{\beta}{3}\right)^2-\frac{1}{2t}\left(t-\frac{\beta}{3}\right)^2+(1-1)r=\frac{2\beta}{3}$$

and the gain of B is

$$\left(\frac{1}{2}+\frac{\beta}{6t}-\left(\frac{1}{2}+\frac{\beta}{6t}\right)\right)r=0.$$

Hence,

$$l_{B,A1}=\lambda\frac{2\beta}{3}\ .$$

We have

$$\begin{aligned}\pi_{A1}&=\frac{1}{2t}\left(t+\frac{\beta}{3}\right)^2+r+l_{A1,2}-l_{B,A1}\\&=\frac{1}{2t}\left(t+\frac{\alpha+\beta}{3}\right)^2+\left(\frac{1}{2}+\frac{\alpha+\beta}{6t}\right)r-\lambda\left(\frac{\alpha(\alpha+2\beta)}{9t}+\frac{2\beta}{3}-\left(\frac{1}{2}-\frac{\alpha+\beta}{6t}\right)r\right),\\\pi_B&=\lambda\frac{2\beta}{3}\ .\end{aligned}$$

(6) When the strategy profile is (N,N):

The resulting downstream outcome is given as

$$v_1=v_2=v+\alpha+\beta;p_1=p_2=t;q_1=q_2=\frac{1}{2};\pi_1=\pi_2=\frac{t}{2}\ .$$

In the negotiation between A1 and 2, with the disagreement outcome of A1's not providing its premium content to 2, the gain of 2 is

$$\frac{t}{2}-\frac{1}{2t}\left(t-\frac{\beta}{3}\right)^2$$

and the gain of A1 is

$$\frac{t}{2}-\frac{1}{2t}\left(t+\frac{\beta}{3}\right)^2+\left(1-\frac{1}{2}-\frac{\beta}{6t}\right)r.$$

Hence,

$$l_{A1,2}=\lambda\left(\frac{\beta}{3}-\frac{\beta^2}{18t}\right)-(1-\lambda)\left(-\frac{\beta}{3}-\frac{\beta^2}{18t}+\left(\frac{1}{2}-\frac{\beta}{6t}\right)r\right).$$

In the negotiation between B and A1, with the disagreement outcome of B's providing its premium content only to 2, the gain of A1 is

$$\frac{t}{2} - \frac{1}{2t}\left(t - \frac{\beta}{3}\right)^2 + (1-1)r$$

and the gain of B is

$$\left(1 - \frac{1}{2} - \frac{\beta}{6t}\right) r.$$

Hence,

$$l_{B,A1} = \lambda\left(\frac{\beta}{3} - \frac{\beta^2}{18t}\right) - (1-\lambda)\left(\frac{1}{2} - \frac{\beta}{6t}\right) r.$$

In the negotiation between B and 2, with the disagreement outcome of B’s providing its premium content only to A1, the gain of 2 is

$$\frac{t}{2} - \frac{1}{2t}\left(t - \frac{\beta}{3}\right)^2$$

and the gain of B is

$$\left(1 - \frac{1}{2} - \frac{\beta}{6t}\right) r.$$

Hence,

$$l_{B,2} = \lambda\left(\frac{\beta}{3} - \frac{\beta^2}{18t}\right) - (1-\lambda)\left(\frac{1}{2} - \frac{\beta}{6t}\right) r.$$

We have

$$\pi_{A1} = \frac{t}{2} + r + l_{A1,2} - l_{B,A1} = \frac{1}{2t}\left(t + \frac{\beta}{3}\right)^2 + r - \lambda\left(\frac{\beta}{3} + \frac{\beta^2}{18t}\right),$$

$$\pi_B = l_{B,A1} + l_{B,2} = 2\left[\lambda\left(\frac{\beta}{3} - \frac{\beta^2}{18t} + \left(\frac{1}{2} - \frac{\beta}{6t}\right) r\right) - \left(\frac{1}{2} - \frac{\beta}{6t}\right) r\right].$$

Summarizing, the game matrix is given below.

| | $E_2$ | $E_{A1}$ | N |
|---|---|---|---|
| E | $\frac{t}{2}$, $\frac{\lambda(\alpha+\beta)(6t-3r-\alpha-\beta)}{18t}+\left(\frac{1}{2}+\frac{\alpha+\beta}{6t}\right)r$ | $\frac{1}{2t}\left(t+\frac{\alpha+\beta}{3}\right)^2+\left(\frac{1}{2}+\frac{\alpha+\beta}{3t}\right)r-\frac{\lambda(\alpha+\beta)(6t+6r+\alpha+\beta)}{18t}$, $\frac{\lambda(\alpha+\beta)(6t+6r+\alpha+\beta)}{18t}+\frac{r}{2}$ | $\frac{1}{2t}\left(t+\frac{\beta}{3}\right)^2+\left(1+\frac{\beta}{6t}\right)r-\lambda\left(\frac{\beta}{3}+\frac{\beta^2}{18t}+\left(\frac{1}{2}+\frac{\beta}{6t}\right)r\right)$, $\frac{(\alpha+\beta)r}{6t}+\lambda\left(\frac{\alpha+\beta}{3}+\frac{\beta^2-\alpha^2-2\alpha\beta}{18t}+\left(1-\frac{\alpha}{6t}\right)r\right)$ |
| N | $\frac{t}{2}+\frac{r}{2}+\lambda\left(\frac{\beta^2}{9t}+\frac{r}{2}\right)$, $\lambda\frac{2\beta}{3}$ | $\frac{1}{2t}\left(t+\frac{\alpha+\beta}{3}\right)^2+\left(\frac{1}{2}+\frac{\alpha+\beta}{6t}\right)r-\lambda\left(\frac{\alpha(\alpha+2\beta)}{9t}+\frac{2\beta}{3}-\left(\frac{1}{2}-\frac{\alpha+\beta}{6t}\right)r\right)$, $\lambda\frac{2\beta}{3}$ | $\frac{1}{2t}\left(t+\frac{\beta}{3}\right)^2+r-\lambda\left(\frac{\beta}{3}+\frac{\beta^2}{18t}\right)$, $2\left[\lambda\left(\frac{\beta}{3}-\frac{\beta^2}{18t}+\left(\frac{1}{2}-\frac{\beta}{6t}\right)r\right)-\left(\frac{1}{2}-\frac{\beta}{6t}\right)r\right]$ |

# References

Anderson, S. and S. Coate (2005), "Market provision of broadcasting: A welfare analysis," *Review of Economic Studies* 72, pp. 947-972.

Armstrong, M. (1999), "Competition in the pay-TV market," *Journal of the Japanese and International Economics* 13, pp. 257-280.

Bester, H. and E. Petrakis (1993), "The incentives for cost reduction in a differentiated industry," *International Journal of Industrial Organization* 11, pp. 519-534.

Collard-Wexler, A., G. Gowrisankaran and R. Lee (2019), ""Nash-in-Nash" bargaining: A microfoundation for applied work," *Journal of Political Economy* 127, pp. 163-195.

D'Annunzio, A. (2017), "Vertical integration in the TV market: Exclusive provision and program quality," *International Journal of Industrial Organization* 53, pp. 114-144.

Dobson, P. and M. Waterson (1996), "Exclusive trading contracts in successive differentiated duopoly," *Southern Economic Journal* 63, pp. 361-377.

Dobson, P. and M. Waterson (2007), "The competition effects of industry-wide vertical price fixing in bilateral oligopoly," *International Journal of Industrial Organization* 25, pp. 935-962.

Gabszewicz, J., D. Laussel and N. Sonnac (2001), "Press advertising and the ascent of the 'pensée unique'?", *European Economic Review* 45, pp. 645-651.

Gabszewicz, J., D. Laussel and N. Sonnac (2002), "Press advertising and the political differentiation of newspapers," *Journal of Public Economic Theory* 4, pp. 249-259.

Gabszewicz, J., D. Laussel and N. Sonnac (2004), "Programming and advertising competition in the broadcasting industry," *Journal of Economics & Management Strategy* 13, pp. 657-669.

Gal-Or, E. and A. Dukes (2003), "Minimum differentiation in commercial media markets," *Journal of Economics & Management Strategy* 12, pp. 291-325.

Ghili, S. (2022), "Network formation and bargaining in vertical markets: The case of narrow networks in health insurance," *Marketing Science* 41, pp. 501-527.

Harbord, D. and M. Ottaviani (2001), "Contracts and competition in the pay-TV market," manuscript.

Hart, O. and J. Tirole (1990), "Vertical integration and market foreclosure," *Brookings Papers on Economic Activity: Microeconomics*, pp. 205-286.

Ho, K. and R. Lee (2019), "Equilibrium provider networks: Bargaining and exclusion in Health care markets," *American Economic Review* 109, pp. 473-522.

Horn, H. and A. Wolinsky (1988), “Bilateral monopolies and incentives for merger,” *Rand Journal of Economics* 19, pp. 408-419.

Liebman, E. (2022), “Bargaining in markets with exclusion: An analysis of health insurance networks,” manuscript.

McAfee, P. and M. Schwartz (1994), “Opportunism in multilateral vertical contracting: Nondiscrimination, exclusivity, and uniformity,” *American Economic Review* 84, pp. 210-230.

Nocke, V. and P. Rey (2018), “Exclusive dealing and vertical integration in interlocking relationships,” *Journal of Economic Theory* 177, pp. 183-221.

O’Brien, D. and G. Shaffer (1992), “Vertical control with bilateral contracts,” *Rand Journal of Economics* 23, pp. 299-308.

Peitz, M. and T. Valletti (2008), “Content and advertising in the media: Pay-tv versus free-to-air,” *International Journal of Industrial Organization* 26, pp. 949-965.

Rey, P. and T. Vergé (2004), “Bilateral control with vertical contracts,” *Rand Journal of Economics* 35, pp. 728-746.

Rey, P. and T. Vergé (2019), “Secret contracting in multilateral relations,” manuscript.

Rogerson, W. (2012), “Vertical mergers in the video programming and distribution industry: The case of Comcast-NBCU,” manuscript.

Stennek, J. (2014), “Exclusive quality − Why exclusive distribution may benefit the TV-viewers,” *Information Economics and Policy* 26, pp. 42-57.

Weeds, H. (2014), “Advertising and the distribution of content,” manuscript.

Weeds, H. (2016), “TV wars: Exclusive content and platform competition in pay TV,” *Economic Journal* 126, pp. 1600-1633.